\documentclass[aps,prl,print,twocolumn,groupedaddress]{revtex4-2}
\usepackage{graphicx}
\usepackage{amssymb}
\usepackage{amsfonts}
\usepackage{amsmath}
\usepackage{natbib}
\usepackage{hyperref}
\usepackage{color}
\usepackage{bm}
\usepackage{mathrsfs}
\usepackage{amsthm}

\begin{document}

\title{Observation of edge solitons and transitions between them in a trimer circuit lattice}
\author{Rujiang Li$^{1}$}
\thanks{Corresponding author: {rujiangli@xidian.edu.cn}}
\author{Xiangyu Kong$^{1}$}
\author{Wencai Wang$^{1}$}
\author{Yixi Wang$^{1}$}
\author{Yongtao Jia$^{1}$}
\author{Huibin Tao$^{2}$}
\thanks{Corresponding author: {coldfire2000@mail.xjtu.edu.cn}}
\author{Pengfei Li$^{3,4}$}
\author{Ying Liu$^{1}$}
\author{Boris A. Malomed$^{5}$}
\thanks{Corresponding author: {malomed@tauex.tau.ac.il}}

\affiliation{$^1$National Key Laboratory of Radar Detection and Sensing, School of Electronic
Engineering, Xidian University, Xi'an 710071, China}

\affiliation{$^2$School of Software Engineering, Xi'an Jiaotong University,
Xi'an, China}

\affiliation{$^3$Department of Physics, Taiyuan Normal University, Jinzhong, 030619,
China}

\affiliation{$^4$Institute of Computational and Applied Physics, Taiyuan Normal University,
Jinzhong, 030619, Shanxi, China}

\affiliation{$^5$Instituto de Alta Investigaci\'{o}n, Universidad de Tarapac\'{a},
Casilla 7D, Arica, Chile}

\begin{abstract}
In nonlinear topological systems, edge solitons either originate from linear topological edge modes or emerge as nonlinearity-induced localized states without topological protection. While electric circuits (ECs) provide a platform for realizing various types of topological insulators, observation of edge solitons and transitions between them in EC lattices remains a challenging problem. Here, we realize quench dynamics in nonlinear ECs to experimentally demonstrate both topologically nontrivial and trivial edge solitons in a trimer EC lattice and transitions between them. In the weakly nonlinear regime, we observe two types of topologically nontrivial edge solitons that originate from the corresponding linear topological edge states, characterized by the presence of mutually antisymmetric or symmetric peaks at two edge sites. Under strong nonlinearity, topologically trivial edge solitons with antisymmetric, symmetric, and asymmetric internal structures are discovered. The work suggests possibilities for exploring sophisticated nonlinear states and transitions between them in nonlinear topological systems.
\end{abstract}

\maketitle

\noindent \textbf{Introduction}

\noindent Topological insulators (TIs) are media that act as conventional insulators
in the bulk but maintain conductivity on their surfaces, provided by
topologically protected edge states \cite{TI1,TI2,RMP82-3045,RMP83-1057,QF3-14,QF3-21}.
Counterparts of TIs have been realized across diverse physical platforms,
including acoustic and mechanical ones \cite%
{NRP1-281,NRM7-974,NRP5-483,RMP96-021002,QF3-26}, bosonic condensates in ultracold
gases \cite{RMP91-015005}, and photonics \cite%
{Segev,RMP91-015006,LSA9-130,QF1-10}, the immunity of
topological edge states to local deformations and disorder being crucial for
promising potential applications. Further, the
interplay between the topological structure of optical media and their intrinsic
nonlinearity \cite{APR7-021306, NP20-905} leads to the creation of edge solitons,
which originate from the linear topological edge states and inherit their topological protection \cite%
{PRA90-023813,PRL117-143901,PRA94-021801,optica3-1228,
PRL119-253904,PRL123-254103,PRB106-195423,ncommun11-1902, LPR13-1900223,
PRB102-115411,nphys18-678,OL45-6466,PRL121-163901,NJP22-103058,
PRX11-041057,PRL128-093901,nphys17-995,LSA,LSA2,PRB108-184301}.
In addition to the
topologically nontrivial edge solitons, conventional ones, which are nonlinearity-induced
localized states at the edge of a bulk optical waveguide, have also been
discovered in nonlinear photonic topological insulators \cite{PRA94-021801,nphys17-995,
PRB104-235420,PRB102-115411,nphys18-678}.
The conventional edge solitons are considered as topologically trivial states, since they do not originate from linear topological edge modes.
However, historically, they have been refereed to as surface solitons and regarded as important members of the soliton family \cite{PR463-1,RPP75-086401,RMP83-247}.
It is relevant to mention that trivial edge solitons emerge with the power exceeding a certain finite threshold
value \cite{Segev-063901,PRB102-115411,PR463-1,PRL128-093901}.

Electric circuits (ECs) have been widely used as a versatile platform for
simulations of a great variety of nonlinear modes which are known in other
areas of physics \cite{English}. In particular, this platform was recently
proposed as a means for emulating various types of TIs \cite%
{nphys14-875,CP1-39,AP399-289, PRResearch3-023056,PR1093-1,arxiv-lee,PRB107-245114,FP20-44602}. In this
context, owing to the broad flexibility in constructing EC lattices and
employing site-resolved, phase-resolved, time-resolved, and
frequency-resolved measurement techniques, ECs have demonstrated their
relevance for exploring multidimensional \cite%
{PRRsearch4-033203,ncommun11-2356}, higher-order \cite{nphys14-925,LSA9-145}%
, non-Hermitian \cite{nphys16-747,research2021-5608038, ncommun12-7201,PRL133-136602,FP20-44204},
non-Abelian \cite{nelectron5-635,nature594-195}, and non-Euclidean TIs \cite%
{ncommun13-2937,ncommun14-1083}. While some work on nonlinear topological
ECs has been reported \cite{nelectron1-178,ncommun10-1102,PRL123-053902,PNAS118-e2106411118,PRResearch5-L012041,FP18-33311,
arxiv2411.07522},
these studies typically rely on driving the lattices with continuous voltage sources. This approach works well in simple systems, but
becomes cumbersome when applied to the study of nonlinear topological lattices which exhibit a broad variety of solitons,
as it necessitates additional considerations such as taking care of bistability \cite{PRL119-253904,PRL121-163901,nphys18-678,arxiv-prb}.
Due to the lack of a suitable realization scheme that can selectively excite the desired nonlinear states, the observation of edge solitons in EC lattices remains a significant challenge -- particularly, transitions between topologically nontrivial and trivial edge solitons, as well as transitions between different types of topological states \cite{Japan}.

In this work, we have experimentally implemented a nonlinear trimer EC
lattice with a linear topological structure, observing both
topologically nontrivial and trivial edge solitons, using the technique of quench dynamics.
We measure the time evolution of site voltages
initiated by the out-of-phase, in-phase, and single-site excitations, while tuning the EC
nonlinearity by the excitation voltage.
For the out-of-phase and in-phase excitations, we consider both
weak and strong nonlinearity. Under the weak nonlinearity,
the excitation creates topologically nontrivial edge solitons originating from the linear topological edge states.
In contrast, under the action of strong nonlinearity, the
initial voltage distribution leads to the creation of conventional edge
solitons, thus indicating that a transition from
topologically nontrivial to trivial
edge solitons occurs with the growth of the nonlinearity
strength. We also explore the cases of both weak and strong nonlinearity for
single-site excitations. For weak nonlinearity, voltage oscillations are
observed due to the overlap of two types of linear
topological edge states. When the excitation voltage exceeds a threshold
value, strong voltage localization around a single edge site occurs, forming
a topologically trivial asymmetric edge soliton.
The discovery of the three types of topologically trivial edge solitons was not reported in the previous study of a nonlinear
trimer photonic array \cite{PRL128-093901}.
This work offers a general approach to the creation of solitons in EC lattices,
which may be also helpful for the complete understanding of exotic self-trapped states and phase diagrams
in more complex nonlinear topological systems.
\newline

\noindent \textbf{Results}

\noindent \textbf{Implementation of a nonlinear trimer EC
lattice and the realization of quench dynamics.} Although the Su-Schrieffer-Heeger lattice is the prototypical model of a topological lattice, we here investigate a one-dimensional trimer lattice with onsite nonlinearity, to highlight the powerful capabilities of quench dynamics. We will show that, in contrast to the nonlinear Su-Schrieffer-Heeger lattice \cite{PRB102-115411}, the nonlinear trimer lattice supports a diverse range of edge solitons, both topologically nontrivial and trivial ones being effectively excited and probed through phase-resolved techniques.

\begin{figure*}[tbp]
\includegraphics[width=17.7cm]{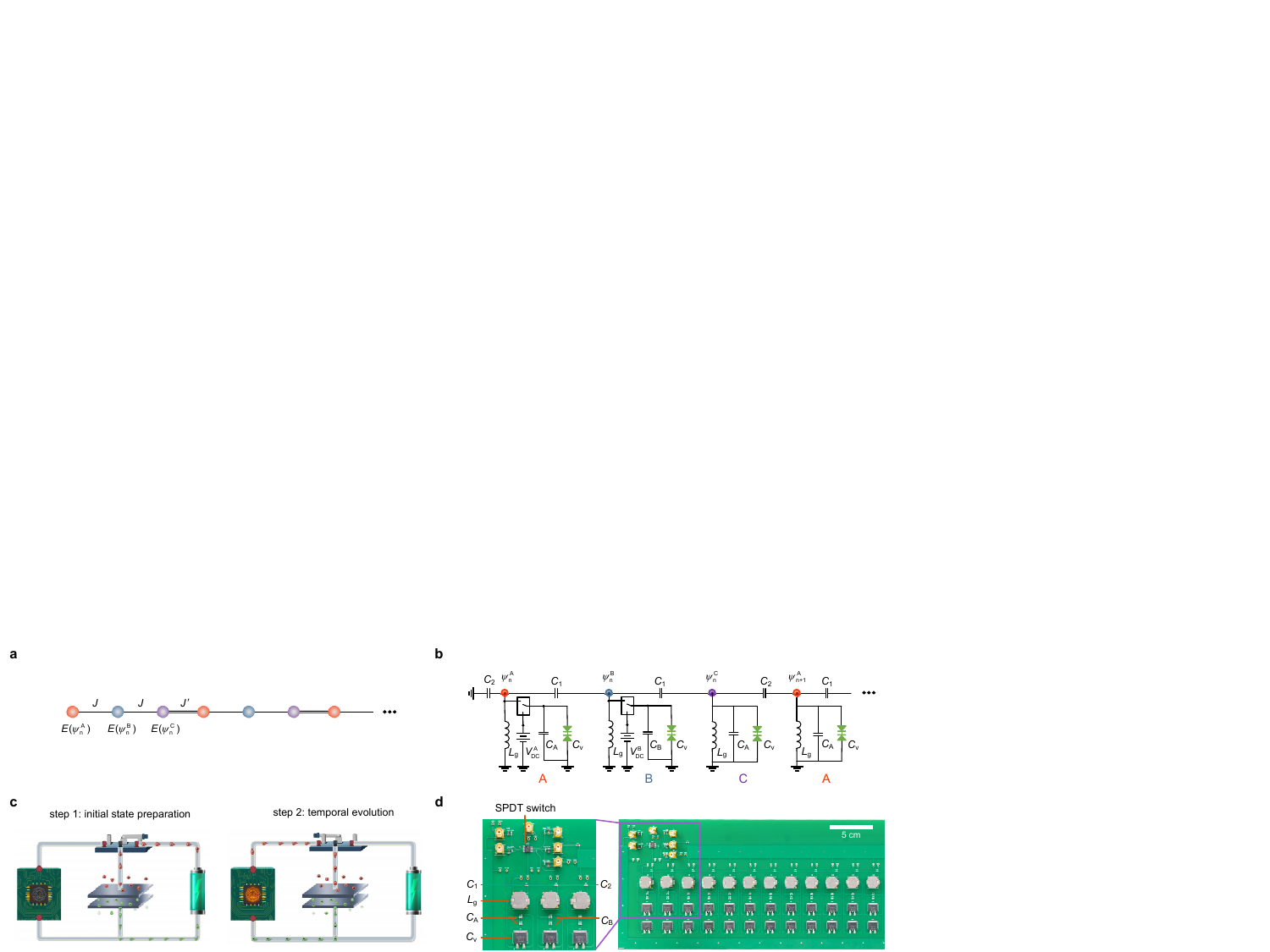}
\caption{\textbf{The implementation of the nonlinear trimer EC
lattice and the realization of quench dynamics.} \textbf{%
a:} The schematic of the nonlinear trimer lattice, where $J$ and $J^{\prime
} $ represent the intra- and inter-cell hopping amplitudes, respectively,
and onsite energy $E$ at each site depends on the wave function of that
site, $\protect\psi _{n}^{\text{A,B,C}}$. \textbf{b:} The circuit
implementation of the nonlinear trimer lattice, where the node voltages
correspond to the wave functions at the lattice sites, capacitors $C_{1}$
and $C_{2}$ emulate the intra- and inter-cell hopping amplitudes,
respectively. The onsite nonlinearity is provided by the common-cathode diodes,
which exhibit voltage-dependent capacitance $C_{\text{v}}$. In the two
leftmost nonlinear $LC$ oscillators, SPDT switches and DC voltage sources
are employed to implement the quench dynamics. \textbf{c:} The illustration
of the quench dynamics, which involves two steps, \textit{viz.}, the
preparation of the initial state and its subsequent time evolution. The charging of
capacitors and diodes corresponds to the
preparation of the initial state, while the discharging represents the
evolution in the EC lattice.
\textbf{d:} The experimental sample of the nonlinear trimer EC lattice. The inset shows an
enlarged fragment with typical circuit components. Experimentally, we use two common-cathode diodes
connected in parallel to provide the capacitance $C_{\text{v}}$.}
\label{fig1}
\end{figure*}

As shown in Fig. %
\ref{fig1}a, the nonlinear trimer lattice is governed by the following
dynamical equations:%
\begin{eqnarray}
\mathrm{i}\frac{\mathrm{d}\psi _{n}^{\text{A}}}{\mathrm{d} t} &=&E\left( \psi _{n}^{\text{A}}\right)
\psi _{n}^{\text{A}}+J^{\prime }\psi _{n-1}^{\text{C}}+J\psi _{n}^{\text{B}},
\label{eq1} \\
\mathrm{i}\frac{\mathrm{d} \psi _{n}^{\text{B}}}{\mathrm{d} t} &=&E\left( \psi _{n}^{\text{B}}\right)
\psi _{n}^{\text{B}}+J\psi _{n}^{\text{A}}+J\psi _{n}^{\text{C}},
\label{eq2} \\
\mathrm{i}\frac{\mathrm{d} \psi _{n}^{\text{C}}}{\mathrm{d} t} &=&E\left( \psi _{n}^{\text{C}}\right)
\psi _{n}^{\text{C}}+J\psi _{n}^{\text{B}}+J^{\prime }\psi _{n+1}^{\text{A}}.
\label{eq3}
\end{eqnarray}%
Here, $J$ and $J^{\prime }$ are the intra- and inter-cell hopping
amplitudes, respectively, and $E\left( \psi _{n}^{\text{A,B,C}}\right) $ are
the onsite energies which depend on the respective wave functions.
To analyze the properties of the states in the lattice, a common approach is to study the quench dynamics by defining
an initial state $\psi_{n}^{\text{A,B,C}}(t=0)$ and exploring its time evolution \cite{optica3-1228,NJP22-103058,PRB106-195423}.
Note that the quench dynamics defined here refers to the general dynamical behavior of a prepared initial state, which is distinct from
the oscillation quenching observed in systems of coupled nonlinear oscillators, specifically including the
oscillation-death and amplitude-death
behaviors \cite{ncommun14-5515,as6-1900771,nanophotonics10-2883,AEM8-2200579,ACS10-147}.
In optical waveguide arrays, the quench dynamics we refer to is typically investigated by replacing the time dimension with
an extra spatial dimension \cite{PRA90-023813,PRL117-143901,PRA94-021801,OL45-6466,PRX11-041057,nphys17-995,PRL128-093901}.

We implement the above model including the quench dynamics using the nonlinear EC lattice shown in
Fig. \ref{fig1}b. Each unit cell includes three nonlinear $LC$ resonators
with inductor $L_{\text{g}}$, capacitors $C_{\text{A}}$ or $C_{\text{B}}$,
and common-cathode diodes $C_{\text{v}}$, with capacitance $C_{\text{v}}=C_{%
\text{L}}/\left( 1+\left\vert v/v_{0}\right\vert \right) ^{M}$, where $C_{%
\text{L}}$, $v_{0}$, and $M$ are constants, and $v$ is the amplitude of the voltage applied
to terminals of the common-cathode diode (the model of this element is
presented in Supplementary Note 1). The nonlinear resonators are coupled via capacitors $C_{1,2}$.
The four types of capacitors satisfy the relation: $C_{\text{A}}-C_{\text{B}}=C_{1}-C_{2}$.
In the two leftmost resonators, single-pole double-throw (SPDT) switches control the charging and discharging of the
respective capacitors $C_{\text{A,B}}$ and diodes $C_{\text{v}}$.

From the EC schematic and illustration of quench dynamics shown in Figs. \ref%
{fig1}b and c, respectively, it is seen that, when the SPDT switches are
connected to the DC voltage sources, capacitors $C_{\text{A,B}}$ and diodes $%
C_{\text{v}}$ are charged to constant voltages $V_{\text{DC}}^{\text{A,B}}$.
The charging operation corresponds to the preparation of the initial state. When the SPDT switches are simultaneously toggled to the circuit nodes,
the instantaneous voltages at the two leftmost circuit nodes are given by $\psi_{1}^{\text{A}}=V_{\text{DC}}^{\text{A}}$ and $\psi_{1}^{\text{B}}=V_{\text{DC}}^{\text{B}}$,
while all other nodes have zero voltage. This moment is recorded as $t=0$. It is worth noting that the phase of the initial state can be adjusted using the DC voltage sources.

At $t > 0$, the charged capacitors and diodes discharge. Under the slowly-varying envelope approximation, which is ensured by
the conditions $C_{1,2} \ll C_{\text{A}} + C_{\text{L}}$
and $C_{\text{L}}-C_{\text{v}} \ll C_{\text{A}} + C_{\text{L}}$ (see the parameters for circuit components in Methods, Sample fabrication and measurement),
the time evolution of the initial voltage distribution is governed by equations
that take the same form as Eqs. (\ref{eq1})-(\ref{eq3}) (for the derivation of the EC equations, see Supplementary Note 1;
for the detailed implementation of the quench dynamics, refer to Methods, Sample Fabrication and Measurement, and Supplementary Note 1).
In contrast to previous work that begins with the Lagrangian \cite{nelectron1-178}, we present an ab initio derivation starting from the Kirchhoff circuit equations.
These time-dependent equations are fundamental for conducting the study of quench dynamics.

From the expressions $J_{1,2} = \frac{C_{1,2}}{2\left(C_{\text{A}} + C_{\text{L}}\right)} \omega_{0}$ with
$\omega _{0}=\frac{1}{\sqrt{L_{\text{g}}\left( C_{\text{A}}+C_{\text{L}}\right) }}$ as the carrier frequency (see Supplementary Note 1 for these definitions),
the intracell and intercell hopping amplitudes $J$ and $J^{\prime}$
are related to the coupling capacitors $C_{1}$ and $C_{2}$, respectively. The onsite energies are expressed as $E\left( \psi_{n}^{\text{A,B,C}} \right) = E_{0} +
g\left( \psi_{n}^{\text{A,B,C}} \right)$, where $E_{0} = \left[ 1 - \frac{C_{1} + C_{3}}{2\left(C_{\text{A}} + C_{\text{L}}\right)} \right] \omega_{0}$ is a constant term,
and $g\left( \psi_{n}^{\text{A,B,C}} \right) = \frac{C_{\text{L}} - C_{\text{v}}\left(\psi_{n}^{\text{A,B,C}}\right)}{2\left(C_{\text{A}} + C_{\text{L}}\right)} \omega_{0}$
represents the voltage-dependent terms. The wave functions $\psi_{n}^{\text{A,B,C}}$ are defined as the complex amplitudes of the temporal voltages $V_{n}^{\text{A,B,C}} \left(t\right)$,
i.e., $V_{n}^{\text{A,B,C}} \left( t \right)= \frac{1}{2} \psi_{n}^{\text{A,B,C}}\left(t\right) + \text{c.c.}$ Under these correspondences, the resonant frequencies of the EC lattice correspond to the frequency spectra
of the nonlinear trimer lattice.

Figure \ref{fig1}d shows the experimental sample of the nonlinear trimer EC lattice, with
the inset zooming in a fragment containing the typical circuit components (see the sample design and fabrication in Methods, Sample fabrication and measurement).
Note that, in our experiments, we use two common-cathode diodes connected in parallel to provide the desired capacitance $C_{\text{v}}$.
\newline

\noindent \textbf{Frequency spectra and profiles of edge solitons.} In the
linear regime, which is realized for an infinitesimal amplitude of wave
functions when $\psi _{n}^{\text{A,B,C}}$ approach zero, the trimer EC lattice exhibits
three dispersive bands separated by two finite bandgaps under periodic
boundary condition (see Supplementary Figure S5 and relevant discussion in
Supplementary Note 2). The linear trimer lattice being
inversion-symmetric, i.e., $PH\left( k\right) P^{-1}=H\left( -k\right) $,
with Hamiltonian in the reciprocal space $H\left( k\right) $ and inversion
operator $P$, the Zak phase, defined as the integral over the Brillouin
zone, $\mathcal{Z}=\text{i}\int_{\text{BZ}}\left\langle \psi _{k}\right\vert
\frac{\partial}{\partial k}\left\vert \psi _{k}\right\rangle dk$, is quantized, taking
solely values $0$ or $\pi $ (modulo $2\pi $).
When the intracell hopping $J$ is smaller than its intercell
counterpart $J^{\prime }$, the linear trimer lattice is topologically
nontrivial with $\mathcal{Z}=\pi ,~0,$ and $\pi $ for the bottom, middle,
and top bands, respectively.
According to the bulk-boundary correspondence, for the trimer lattice with a finite number of unit cells,
one pair of edge states appear in each topological bandgap
\cite{PRA96-032103, PRA99-013833,PRA103-023503,PRB103-014110,PRB106-085109}.
Due to the inversion symmetry of the trimer EC lattice, the edge states in the bottom topological
gap exhibit the largest magnitudes at the two outermost sites, but with opposite signs,
resulting in antisymmetric internal structures (see discussion in
Supplementary Note 2). In contrast, the edge states in the top topological bandgap also show the largest magnitudes
at the two outermost sites, but with the same sign, leading to symmetric internal structures.

In the nonlinear regime, to give a better physical insight into the existence of edge solitons, we first move to the continuum limit,
where $\psi_{n}^{\text{A,B,C}} \rightarrow \psi_{\text{A,B,C}}(x)$, with $x$ representing the continuum limit index.
We investigate the evolution of solution trajectories with respect to $x$ through a phase-space analysis of nonlinear
systems \cite{LPR13-1900223, nelectron1-178, ACS4-1974}. This approach allows us to predict that edge solitons exist
in a semi-infinite trimer lattice under both weak and strong nonlinearity (for a detailed discussion on equilibrium points and solution trajectories,
see Supplementary Figures S7 and S8, and Supplementary Note 3). In the weakly nonlinear regime, we identify two types of solutions.
One type exhibits $\psi_{\text{A}} \left(x\right) \approx
-\psi_{\text{B}} \left(x\right)$ and $\psi_{\text{C}} \left(x\right) \approx 0$, which corresponds to the antisymmetric internal structure.
The other type shows $\psi_{\text{A}} \approx \psi_{\text{B}}$
and $\psi_{\text{C}} \approx 0$, which corresponds to the symmetric internal structure. Both types of solutions decay monotonically as one
moves away from the lattice boundary. In contrast, in the strongly nonlinear regime, assuming $\psi_{\text{C}} \approx 0$, we predict the existence
of two types of solutions with nearly antisymmetric and symmetric internal structures, respectively, as well as a class of solutions characterized
by strongly asymmetric profiles, where $\psi_{\text{A}} \gg \psi_{\text{B}}$.

\begin{figure*}[tbp]
\includegraphics[width=18cm]{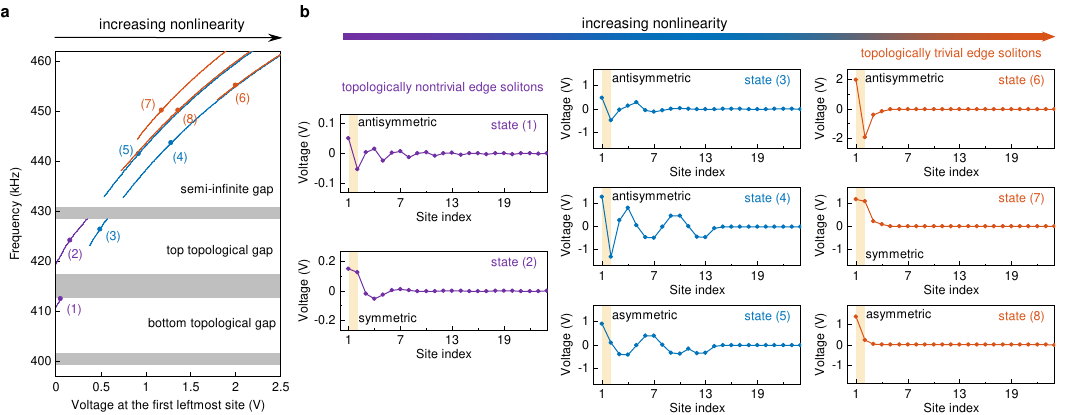}
\caption{\textbf{Frequency spectra and typical profiles of edge solitons.}
\textbf{a:} The frequency spectra as a function of the voltage at the first
leftmost site $\protect\psi _{1}^{\text{A}}$.
The shaded regions represent the underlying spectra of the linear bulk bands. The three bandgaps, \textit{viz.},
the semi-infinite gap, top topological gap,
and bottom topological gap, are defined with respect to the linear bulk spectra.
In the case of weak nonlinearity, two topologically nontrivial edge solitons (purple branches)
originate from the linear
topological edge states. As the voltage increases, distinct families of
edge solitons emerge (blue branches). When nonlinearity is
sufficiently strong, the topologically trivial edge solitons emerge (orange branches). \textbf{b:} Profiles
of the edge solitons which are indicated by numbers (1)-(8) in (\textbf{a}).
The edge solitons are labeled as ``antisymmetric'', ``symmetric'', or ``asymmetric'' based on the internal structures 
at the two leftmost sites (indicated by yellow bars).
The states in the first and third columns correspond to the topologically nontrivial and trivial edge solitons, respectively.}
\label{fig2}
\end{figure*}

We then numerically solve Eqs. (\ref{eq1})-(\ref{eq3}) and the calculated voltage-dependent frequency
spectra are displayed in Fig. \ref{fig2}a (see the numerical method in Methods, Theoretical calculations).
The solid curves represent the edge solitons. For comparison, the underlying frequency spectra of the linear bulk bands
are denoted as the shaded regions. The three bandgaps, \textit{viz.}, the semi-infinite gap, top topological gap,
and bottom topological gap, are defined with respect to the linear bulk spectra.
We use the voltage at the leftmost site, $\psi _{1}^{%
\text{A}}$, to characterize the strength of the EC nonlinearity. Without the
loss of generality, we focus on the edge solitons located at the left edge
of the trimer EC lattice.

Under weak nonlinearity, as indicated by the two
purple curves commencing from $\psi _{1}^{\text{A}}=0$,
the topologically nontrivial edge solitons originate from the linear topological edge states.
Inheriting the internal structures from the parent linear states, from the voltage distributions,
the nontrivial edge solitons in the bottom and top topological gaps exhibit the
nearly antisymmetric and symmetric peaks at the two leftmost sites, respectively
[see profiles of states (1) and (2) in Fig. \ref{fig2}b].
For comparison with other types of edge solitons that will be introduced later, we refer to states (1) and (2) as
antisymmetric and symmetric nontrivial edge solitons, respectively. These two types of nontrivial edge solitons are
coincident with the solutions under weak nonlinearity in the continuum limit.

As $\psi _{1}^{\text{A}}$
increases, the nontrivial edge solitons, i.e., states (1) and (2), become delocalized, entering the
linear bulk bands \cite{PRL128-093901} (see Supplementary Figures S9-S11
for complete frequency spectra and Supplementary Note 4 for detailed discussion).
However, as indicated by the blue curves, families of edge solitons emerge in the top topological bandgap
and semi-infinite one [see states (3)-(5) and the corresponding branches],
being completely separated from the branches of states (1) and (2) \cite{PRL128-093901}.
These edge solitons are weakly localized, especially when they are in the vicinity of the band edges of the linear bulk states.
States (3) and (4) exhibit the antisymmetric internal structures from their voltage distributions,
while state (3) exists only in a narrow region of the top topological gap, becoming delocalized upon entering the
linear bulk band. Although the branch for state (5) is close to that of the
symmetric nontrivial edge solitons [state (2)], it displays a strongly
asymmetric profile at the two leftmost sites.
Based on the internal structures, states (3), (4), and (5) in Fig. \ref{fig2}b are labeled as ``antisymmetric'', ``symmetric'',
and ``asymmetric'', respectively.

When the nonlinearity becomes
sufficiently strong, self-sustained states emerge at the edge of the trimer
lattice, where the nonlinearity dominates over the lattice couplings.
From the voltage distributions, states
(6)-(8) in Fig. \ref{fig2}\text{b} show typical profiles of the three types of edge solitons, featuring antisymmetric (staggered), nearly symmetric
(unstaggered), and asymmetric (strongly confined) internal structures, respectively.
These edge solitons are conventional self-trapped topologically trivial modes \cite{PR463-1,RPP75-086401,RMP83-247,Segev-063901}.
We refer to states (6)-(8) as antisymmetric, symmetric, and asymmetric trivial edge solitons, respectively.
Due to their fundamentally different physical origins, the branches corresponding to these topologically trivial edge solitons
are completely isolated from those of the nontrivial edge solitons \cite{PRB102-115411}.
Note that the existence of these nontrivial edge solitons validates the prediction produced by the phase-space analysis.

Among all the edge solitons shown in Fig. \ref{fig2}b, the topologically trivial ones (6)-(8), which exist in the
semi-infinite gap, exhibit the strongest exponential localization with the
voltage distribution concentrated almost entirely at the two leftmost sites
[or solely at the first leftmost site, for state (8)]. On the other hand,
the topologically nontrivial edge solitons labeled (1)-(2) in Fig. \ref{fig2}b feature
slowly decaying oscillatory tails.
These features indicate the existence of a nonlinearity-induced transition from topologically
nontrivial edge solitons to trivial ones \cite{PRB104-235420}.
Additionally, we note that the frequency spectra shown in Fig. \ref{fig2}a are concentrated around the carrier frequency $\omega_{0}$.
Therefore, the slowly-varying envelope approximation, used to derive equations similar to Eqs. (\ref{eq1})-(\ref{eq3}) is valid,
provided that the voltages are not excessively large.
\newline

\noindent \textbf{Observation of edge solitons and transitions between them.}
To experimentally investigate the topologically nontrivial and trivial edge
solitons and the transitions between them, we implement quench dynamics
in the nonlinear trimer EC lattice. Following the scheme shown in Figs. \ref%
{fig1}b and c, we create the input with $\psi _{n}^{\text{A,B,C}}\left(
t=0\right) $ to match the expected internal structure of the edge solitons.
Specifically, for the antisymmetric and symmetric cases, we set $V_{\text{DC}}^{\text{A}} = -V_{\text{DC}}^{\text{B}} = \psi_{0}$
and $V_{\text{DC}}^{\text{A}} = V_{\text{DC}}^{\text{B}} = \psi_{0}$, respectively. This indicates that the initial voltages are out of phase
and in phase, respectively. For the asymmetric edge solitons, we set $V_{\text{DC}}^{\text{A}} = \psi_{0}$ and $V_{\text{DC}}^{\text{B}} = 0$,
meaning that only the first leftmost site is excited. Although multiple edge solitons are present in Fig. \ref{fig2}, only the one that matches the
initial input distribution and is dynamically stable can be effectively excited.
Under an input excitation, we record the voltage distributions in the EC lattice at various time intervals. If a steady state is achieved,
the voltage distribution will correspond to an eigenstate. This allows us to identify which of the independent edge solitons is being excited.
Similar to optical waveguide arrays, where nonlinearity is influenced by optical power \cite{PRL128-093901},
the nonlinearity in the EC lattice is governed by the value of $\psi_0$.

\begin{figure*}[tbp]
\includegraphics[width=8.6cm]{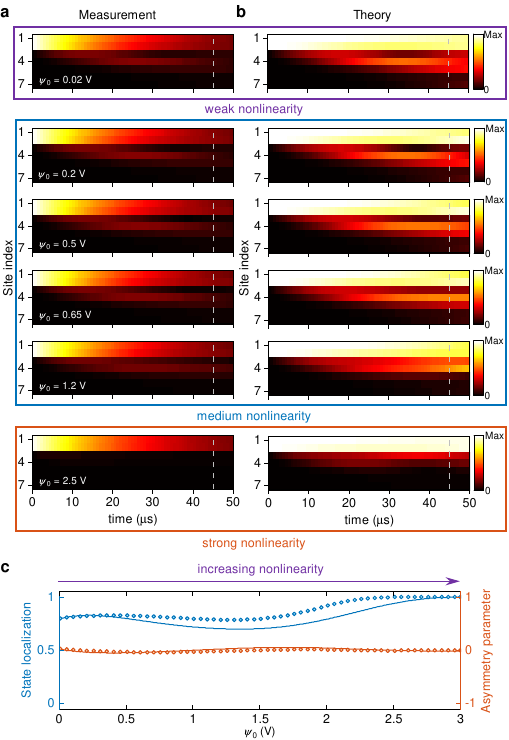}
\caption{\textbf{The quench dynamics initiated by the out-of-phase excitations.}
\textbf{a and b:} The experimentally observed and
theoretically predicted evolution initiated by different initial voltage
excitations. \textbf{c:} The state localization and asymmetry parameter, defined
as per Eqs. (\protect\ref{S2}) and (\protect\ref{Theta}), respectively, as
extracted from the voltage distributions at $t=45~\mathrm{\protect\mu }\text{%
s}$ [this time moment is indicated by the dashed lines in (\textbf{a})-(\textbf{b})].
The continuous curves and chains of circles denote the theoretical and
experimental results, respectively. }
\label{fig3}
\end{figure*}

In the case of out-of-phase excitations, we present the experimentally measured and theoretically calculated temporal evolution
of the site voltages, represented by the magnitudes of the complex amplitudes, $\left\vert \psi_{n}^{\text{A,B,C}} \right\vert$,
in Figs. \ref{fig3}a and b, respectively (see the method for theoretical calculations in Methods, Theoretical calculations).
For $\psi _{0}=0.02~\text{V}$, i.e., under weak nonlinearity, the stability of the antisymmetric
nontrivial edge solitons (see Supplementary Note 4 for the stability
analysis) and substantial overlap between the initial voltage distribution
and state (1) shown in Fig. \ref{fig2} enable the system to achieve a steady
state, before the voltages decay to undetectable levels due to the
dissipative losses from the circuit components. Thus, we observe the formation
of the antisymmetric nontrivial edge soliton.
For $\psi _{0}=2.5~\text{V}$ (strong nonlinearity),
a steady state is also achieved and we observe the formation of state (6) shown in Fig. %
\ref{fig2}, which represents the antisymmetric trivial
edge soliton with strong localization. This outcome results from the nearly
perfect overlap between the initial voltage distribution and established
soliton profiles, along with the stability of these solitons across a broad
voltage range.

For medium nonlinearity strengths, corresponding to $\psi
_{0}=0.2$, 0.5, 0.65, and $1.2~\text{V}$,
steady states are not achieved under the excitations, although the two leftmost sites still exhibit
the highest voltage. When the excitation voltage corresponds to a
delocalized state or a strongly unstable edge soliton, the voltage still
localizes at these two sites due to limited measurement time (See
Supplementary Figure S12 and
Supplementary Note 5). If the excitation voltage corresponds to a weakly
unstable edge soliton, the initial voltage may excite it, but the localized
voltage eventually collapses.

To compare with the above results, we also study a trimer EC lattice with $%
J>J^{\prime }$, which, as mentioned above, is topologically trivial in the linear limit.
We observe voltage diffraction accompanied by local oscillations across the entire range of excitation voltages,
indicating that steady states are never established under varying excitations. This observation confirms the absence of
both topologically nontrivial and trivial edge solitons with an antisymmetric internal structure (see Supplementary Figure S14
and Supplementary Note 6).
Note that the trivial edge solitons do not exist in this case
because the circuit lattices exhibit saturable nonlinearity, and the nonlinearity strength is insufficient to
induce mode localization.

To characterize the excitation of antisymmetric edge solitons
during quench dynamics, we define the state localization $S_{2}$ as follows:
\begin{equation}
S_{2}=\frac{|\psi _{1}^{\text{A}}|^{2}+|\psi _{1}^{\text{B}}|^{2}}{%
\sum\limits_{n}\left( |\psi _{n}^{\text{A}}|^{2}+|\psi _{n}^{\text{B}%
}|^{2}+|\psi _{n}^{\text{C}}|^{2}\right) }.  \label{S2}
\end{equation}%
It measures the voltages of the two leftmost sites relative to the total
voltage across all EC nodes. Additionally, we introduce the asymmetry
parameter $\Theta $ to quantify the imbalance between the voltages at these
two sites:
\begin{equation}
\Theta =\frac{|\psi _{1}^{\text{A}}|-|\psi _{1}^{\text{B}}|}{|\psi _{1}^{%
\text{A}}|+|\psi _{1}^{\text{B}}|}.  \label{Theta}
\end{equation}%
Theoretically, the state localization and asymmetry parameter should be evaluated after a sufficiently long evolution time.
Due to the inevitable circuit dissipations in experiments, we extract the voltage distributions at $t = 45~\mathrm{\mu s}$,
with the corresponding results for the state localization and asymmetry parameter shown in Fig. \ref{fig3}c.
We will demonstrate below that the selection of this time moment is appropriate for
observing the nonlinearity-induced transition between the topologically nontrivial and trivial edge solitons.

The continuous curves and chains of circles in Fig. \ref{fig3}c represent the theoretical and experimental results, respectively.
Although the circuit dissipation causes deviation in the specific values of the experimental results, the overall trends well
agree with the theoretical predictions.
The state localization is relatively high near $\psi _{0}=0$ and at $\psi _{0}>2$, as the
topologically nontrivial and trivial edge solitons are formed under the weak and
strong nonlinearity, respectively. Notably, the trivial edge solitons
[such as state (6) in Fig. \ref{fig2}] are, generally, localized stronger
than the nontrivial ones [such as state (2)], resulting in $S_{2}\approx 1$ for
$\psi _{0}>2.5~\text{V}$. This fact
indicates the above-mentioned nonlinearity-induced transition from the weakly
nonlinear nontrivial edge solitons to the trivial ones supported
by strong nonlinearity \cite{PRB104-235420}.
Note that the same conclusion can be reached by
evaluating the inverse participation ratios (IPRs) of the voltage distributions as nonlinearity increases
(see Supplementary Figure S17 and Supplementary Note 7 for the results of IPRs).
The IPR is defined as follows:
\begin{equation}
\text{IPR} = \frac{\sum\limits_{n} \left( \left\vert \psi _{n}^{\text{A}} \right\vert^{4}
+ \left\vert \psi _{n}^{\text{B}} \right\vert^{4}
+ \left\vert \psi _{n}^{\text{C}} \right\vert^{4} \right)}
{\left[ \sum\limits_{n} \left( \left\vert \psi _{n}^{\text{A}} \right\vert^{2}
+ \left\vert \psi _{n}^{\text{B}} \right\vert^{2}
+ \left\vert \psi _{n}^{\text{C}} \right\vert^{2} \right) \right]^{2}}.
\label{IPR}
\end{equation}
Besides, throughout this transition, the antisymmetric internal structure of the edge
soliton is preserved, with the asymmetry parameter $\Theta $ remaining close to
zero, as shown in Fig. \ref{fig3}c. Phase measurements further support this observation
(see Supplementary Figure S18 and Supplementary Note 8 for the phase measurement of the edge solitons).

\begin{figure*}[tbp]
\includegraphics[width=8.6cm]{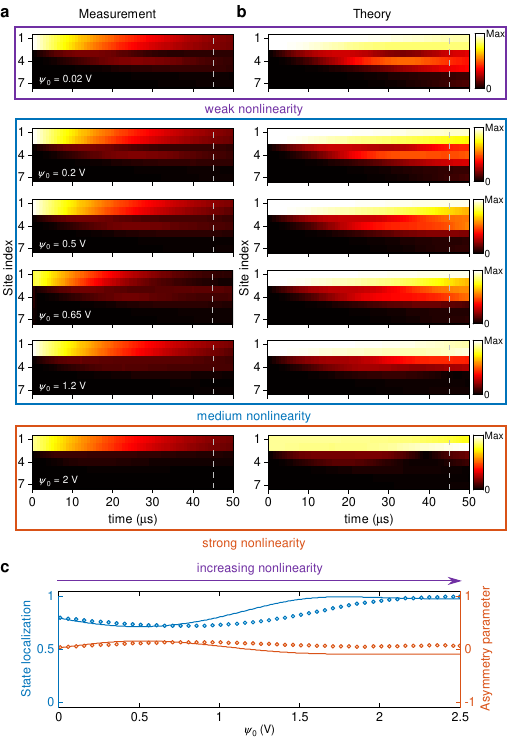}
\caption{\textbf{The quench dynamics initiated by the in-phase excitations.}
\textbf{a and b:} The experimentally observed and
theoretically predicted evolution initiated by different initial voltage
excitations. \textbf{c:} The state localization and asymmetry parameter, defined
as per Eqs. (\protect\ref{S2}) and (\protect\ref{Theta}), respectively, as
extracted from the voltage distributions at $t=45~\mathrm{\protect\mu }\text{%
s}$ [this time moment is indicated by the dashed lines in (\textbf{a})-(\textbf{b%
})]. The continuous curves and chains of circles denote the theoretical and
experimental results, respectively.}
\label{fig4}
\end{figure*}

Similarly, we investigate the evolution of site voltages initiated by in-phase excitations.
As shown in Figs. \ref{fig4}a and b, the formation of
the symmetric nontrivial edge soliton [state (2) in Fig. \ref{fig2}] is
observed in the case of weak nonlinearity with $\psi _{0}=0.02~\text{V}$,
again attributed to the stability of the topologically nontrivial edge soliton and
significant overlap of the excitation distribution with the soliton profile (see Supplementary Figure S9
and Supplementary Note 4). In the case of a medium nonlinearity strength, the two leftmost sites
display the highest voltage due to the limited measurement time. For a
longer evolution time, voltage diffraction or oscillations occur, if the
excitation corresponds to a delocalized state or unstable edge soliton (see
Supplementary Figure S10 and
Supplementary Note 4). Even for $\psi _{0}=2~\text{V}$, in the regime of
strong nonlinearity, the topologically trivial edge soliton [state (7) in Fig. \ref%
{fig2}] can only be observed briefly, before evolving into the asymmetric
one [state (8)] due to the instability.
Note that the voltage values in the regime of the strong nonlinearity differs from those in Figs. 3a-b,
as the symmetric trivial edge solitons can exist under smaller nonlinearities compared to the antisymmetric ones
(see Fig. \ref{fig2}).
For the trimer EC lattice with $J>J^{\prime
}$, voltage diffraction is observed throughout the entire parameter range,
indicating the absence of both symmetric nontrivial and symmetric trivial edge solitons (see Supplementary Figure
S15 and Supplementary Note 6).

We again evaluate the state localization and asymmetry parameter using the voltage distributions at $t = 45~\mathrm{\mu s}$.
In Fig. \ref{fig4}c, the enhanced state localization indicates a transition from weakly nonlinear nontrivial edge solitons to strongly nonlinear trivial ones,
and the nearly zero values of the asymmetry parameter imply the preservation of the symmetric internal structure of the edge solitons
(also see Supplementary Figure S18 and Supplementary Note 8 for the phase measurements).
The discrepancies between the experimental results (represented by chains of circles) and their theoretical counterparts (shown as
continuous curves), particularly in the medium-nonlinearity regime, are explained by effects of the circuit dissipation in the experiment.

\begin{figure*}[tbp]
\includegraphics[width=8.6cm]{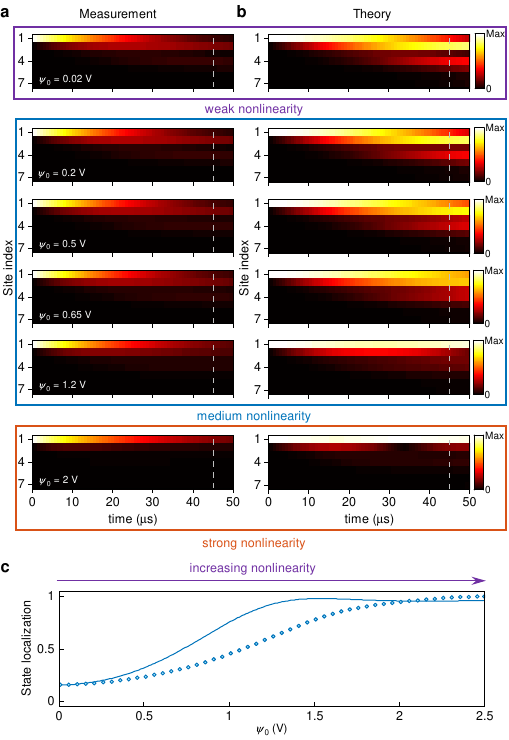}
\caption{\textbf{The quench dynamics initiated by the single-site excitations.}
\textbf{a and b:} The experimentally observed and theoretically predicted
evolution initiated by different initial voltage excitations. \textbf{c:}
The state localization parameter defined as per Eq. (\protect\ref%
{S1}), as extracted from the voltage
distributions at $t=45~\mathrm{\protect\mu }\text{s}$ [this time moment is
indicated by the dashed lines in (\textbf{a})-(\textbf{b})]. The continuous
curve and chain of circles denote the theoretical and experimental
results, respectively. }
\label{fig5}
\end{figure*}

To further clarify the differences between the topologically nontrivial and trivial
edge solitons, we study the quench dynamics initiated by the single-site
excitations applied exclusively to the first leftmost site. In Figs. \ref{fig5}a
and b, one observes voltage oscillations between the two leftmost sites in
the case of weak nonlinearity. These oscillations are induced by the overlap
of the excitation with the linear antisymmetric and symmetric topological
edge states (see Supplementary Figure S10 and Supplementary Note 4).
In the case of medium nonlinearity
strength, the asymmetric edge soliton [state (5) in Fig. \ref{fig2}] does
not form due to its strong instability and weak localization (see Supplementary Figure S10
and Supplementary Note 4).
However, when the nonlinearity is sufficiently strong, such as when $\psi_{0} = 2~\text{V}$, the steady state is achieved and we
observe the formation of the asymmetric trivial edge soliton [state (8)]. Note that we use the same voltage value for the regime of strong nonlinearity
as in Figs. \ref{fig4}a-b, since the symmetric and asymmetric trivial edge solitons require comparable nonlinearity strengths
for their existence (see Fig. \ref{fig2}).

This observation is further verified by a modified definition of the localization parameter,
\begin{equation}
S_{1}=\frac{|\psi _{1}^{\text{A}}|^{2}}{\sum\limits_{n}\left( |\psi _{n}^{%
\text{A}}|^{2}+|\psi _{n}^{\text{B}}|^{2}+|\psi _{n}^{\text{C}}|^{2}\right) }%
,  \label{S1}
\end{equation}%
cf. Eq. (\ref{S2}). As shown in Fig. \ref{fig5}c, when $\psi _{0}$ is close
to zero, the localization is weak due to voltage oscillations. In contrast to
that, we find that $S_{1}\approx 1$ for $\psi _{0}>2~\text{V}$, featuring a
significant enhancement in the state localization.
Here too, the discrepancy between the experimental results (represented by the chain of circles) and
their theoretical counterparts (depicted by the continuous curve) is explained to the effect of the circuit dissipation
in the experiment. In particular, in the regime of the medium nonlinearity, such as the one with $\psi_{0}=1.2~\text{V}$,
the experimental results deviate because the asymmetric
trivial edge soliton is not excited due to its weak localization in this specific regime. The overall trends of the experiential and theoretical results
indicate that there are no weakly-nonlinear asymmetric nontrivial edge solitons, while strongly-nonlinear asymmetric
trivial ones do exist.
Note that asymmetric edge solitons, including both the nontrivial and trivial ones, do not exist in the trimer
EC lattice with $J>J^{\prime}$ (see Supplementary Figure S16 and Supplementary Note 6).
\newline

\noindent \textbf{Discussion}

\noindent The edge solitons and nonlinearity-induced transitions observed in our work are fundamentally distinct from the
nonlinearity-induced topological phase transitions reported previously \cite{nelectron1-178,PRL123-053902}.
In those studies, the EC lattice is topologically trivial in the linear limit.
Under strong excitation, the lattice transitions to a topologically nontrivial phase that supports topologically nontrivial edge states.
In other words, as the excitation strength decreases, these nontrivial edge states reduce to linear trivial bulk states,
rather than linear topological edge states.
In contrast, we develop the technique of quench dynamics and demonstrate the existence of both topologically nontrivial
and trivial edge solitons. Specifically, the nontrivial edge solitons directly arise from the linear topological edge states. Furthermore, we observe the
transitions from topologically nontrivial to trivial edge solitons with enhanced nonlinearity.
These features were not reported in previous works \cite{nelectron1-178,PRL123-053902}.

On the other hand, while both topologically nontrivial and trivial edge solitons are important members of the soliton family in a nonlinear topological lattice,
trivial edge solitons are usually overlooked \cite{PRA90-023813,PRL117-143901,optica3-1228,PRL119-253904,PRL121-163901,
PRL123-254103,NJP22-103058,OL45-6466,ncommun11-1902,PRX11-041057,PRL128-093901,PRB108-184301}.
Notably, Leykam \textit{et al.} proposed edge solitons that originate from the linear topological edge states in a topologically nontrivial lattice,
as well as self-induced nonlinear edge modes in a topologically trivial lattice \cite{PRL117-143901}.
They did not study the conventional trivial edge solitons in a nontrivial lattice.
Similar to our work, Kartashov \textit{et al.} also studied a nonlinear trimer lattice and observed the topologically
nontrivial edge solitons \cite{PRL128-093901}.
However, the topologically trivial edge solitons residing in the semi-infinite gap have not been explored.
The enhanced localization associated with increased nonlinearity, which is an essential signature of the existence of trivial edge solitons
and the transition from topologically nontrivial to trivial edge solitons, was not revealed in their experiments.

Very recently, Sone \textit{et al.} theoretically introduced the concept of the nonlinear Chern number,
which is used to characterize the
nonlinear Chern insulators and nonlinearity-induced topological phase transition \cite{nphys20-1164}.
In contrast to this work, we explore both the regimes of weak
and strong nonlinearity of a lattice which is topologically nontrivial in the linear limit.
In addition to the topologically nontrivial edge solitons that exist under weak nonlinearity
and originate from the linear topological edge states, we also experimentally identify the topologically
trivial edge solitons under strong nonlinearity and establish the nonlinearity-induced transition between the nontrivial and trivial edge solitons.

The coexistence of topologically nontrivial and trivial edge solitons has been experimentally confirmed in two photonic
settings \cite{nphys17-995,nphys18-678}. Compared with these realizations, our platform of nonlinear EC lattices
offers several advantages.
First, we can construct more complex nonlinear topological systems using EC lattices and explore exotic self-trapped states
through the technique of quench dynamics.
Second, the platform of nonlinear EC lattices has the ability of phase-resolved measurements. We additionally provide more experimental evidence of the existence of edge solitons
(see Supplementary Figure S18 and Supplementary Note 8 for the phase measurements of the edge solitons). Finally,
due to the broad availability of circuit components, lattices exhibiting multiple types of nonlinearities can be built \cite{English,arxiv2411.07522}.
For instance, the value of $M$ in the formula
$C_{\text{v}}=C_{\text{L}}/\left( 1+\left\vert v/v_{0}\right\vert \right) ^{M}$ can be adjusted by employing diodes with different parameters.
This flexibility suggests one to explore more complex nonlinearities in EC lattices, in addition to the typical cubic (Kerr) nonlinearity dealt with in conventional
nonlinear optics \cite{APR7-021306,NP20-905,PRL132-213802}.
\newline

\noindent \textbf{Conclusion}

\noindent In this work, we have realized the quench dynamics in the
nonlinear trimer EC lattice with tunable nonlinearity,
observing the time evolution of site voltages initiated by different excitations.
We have experimentally identified the weakly-nonlinear
antisymmetric and symmetric nontrivial edge solitons, as well as the
strongly-nonlinear trivial edge solitons featuring antisymmetric, symmetric,
and asymmetric internal structures.
We have found the nonlinearity-induced transition between the
topologically nontrivial and trivial edge solitons.
Our findings pave the way for further exploration of nonlinear topological physics in EC lattices,
suggesting insights into the interplay of topology and nonlinearity in the
general context.
\newline

\noindent \textbf{Methods}

\noindent \textbf{Sample fabrication and measurement.} To ensure the
observation of the edge solitons, the circuit components should have minimal
parasitic parameters, and their tolerance should be as low as possible. For
this purpose, we utilized capacitors ($C_{\text{A}}=380~\text{pF}$, $C_{%
\text{B}}=760~\text{pF}$, $C_{\text{1}}=180~\text{pF}$, and $C_{\text{2}%
}=560~\text{pF}$) with low ESL and $\pm 1\%$ tolerance. We also employed
inductors with magnetic shielding and low DCR ($L_{\text{g}}=15~\mathrm{\mu }%
\text{H}$), and carefully selected the components using an LCR meter (HIOKI
IM3536). The tolerances for inductance and series resistance are $\pm 1\%$
and $\pm 2\%$, respectively. The average series resistance of the inductors
is approximately $600~\text{m}\Omega $. To characterize the voltage response
of the common-cathode diode (V60DM45C), we measured the $C$-$V$ curves,
obtaining parameter values $C_{\text{L}}=8.62~\text{nF}$, $v_{0}=1.69$, and $%
M=0.31$ by fitting these curves to the phenomenological formula
$C_{\text{v}}=C_{\text{L}}/\left( 1+\left\vert v/v_{0}\right\vert \right) ^{M}$.
We employed standard PCB techniques to fabricate the EC
lattice, ensuring that the inductors are sufficiently spaced to prevent
mutual coupling. The PCB traces were designed with a relatively large width
of $0.75~\text{mm}$ to accommodate high currents, and the layouts were
carefully optimized to minimize parasitic parameters and coupling with other
circuit components.

To implement the quench dynamics, an SPDT switch with two channels (ADG1636)
was used to control the charging and discharging of capacitors $C_{\text{A,B}%
}$ and diodes $C_{\text{v}}$.
To ensure synchronization between the two channels, one specific switch with a small parameter error of $t_{\text{ON/OFF}}$ was selected based on
the measured switching times. Additionally, during the PCB design, the signal paths for the logic control inputs of both channels were deliberately
designed to have equal physical lengths, effectively minimizing propagation delay skew. The capacitor $C_{\text{A}}$ and diode $C_{\text{v}}$
in the leftmost nonlinear $LC$ oscillator were connected to the drain terminal of one channel of the switch, while the capacitor $C_{\text{B}}$ and
diode $C_{\text{v}}$ in the second leftmost nonlinear $LC$ oscillator were connected to the drain terminal of the other channel of the switch.
Two independent DC voltage sources, $V_{\text{DC}}^{\text{A}}$ and $V_{\text{DC}}^{\text{B}}$, were connected to the source terminals of the
two channels through separate SubMiniature version A (SMA) connectors, respectively.
The remaining two source terminals of the switch were each connected to the two leftmost
nodes of the trimer EC lattice. The logic control inputs of the two channels were interconnected and connected to a common external digital signal
generated by an arbitrary function generator (Tektronix AFG31252) via the last SMA connector on the PCB. To minimize coupling noise and spurious signals
induced by the power supplies, including both the positive and negative supplies for the switch, as well as the DC voltage sources used to charge the capacitors
and diodes, we utilized DC power supplies with low ripple and noise. In addition, we implemented extra decoupling capacitors on the PCB.

When we set the external digital signal to a low voltage, the drain terminal of each channel of the switch was connected to one of the source terminals,
allowing the capacitors $C_{\text{A,B}}$ and diodes $C_{\text{v}}$ to charge. Once the capacitors and diodes were charged to their
constant voltages, we switched the digital signal to a high voltage, connecting the drain terminal of each channel to another source terminal. This caused the
capacitors $C_{\text{A,B}}$ and diodes $C_{\text{v}}$ to begin discharging. We recorded the voltage signals at the EC nodes during both the charging and
discharging processes using an oscilloscope that offers high vertical accuracy and low noise floor (Keysight HD304MSO). The post-processing of the recorded
data, including signal capture, filtering, and Fourier analysis to extract both the amplitudes and phases of the complex amplitudes, was carried out.

\noindent \textbf{Theoretical calculations.} To calculate the nonlinear
frequency spectra and profiles of the edge solitons, we solved the
Gross-Pitaevskii equations, i.e., Eqs. (\ref{eq1})-(\ref{eq3}), using the
ansatz $\psi _{n}^{\text{A,B,C}}=\phi _{n}^{\text{A,B,C}}\exp \left(
-i\omega t\right) $, where $\phi _{n}^{\text{A,B,C}}$ is the real-valued
function describing the voltage distribution and $\omega $ is the frequency.
We employed the Newton's method to solve the eigenvalue equation for each $%
\omega $ with appropriate voltage distributions adopted as the initial
guesses \cite{CSF192-116302}. Open boundary conditions were used to truncate the nonlinear EC
lattice. Once we obtained the soliton solution at a given $\omega $,
solutions at other frequencies were obtained iteratively. The stability of
the edge solitons was analyzed using the standard linear-stability technique
and subsequently confirmed through simulations of the time evolution,
based on the Runge-Kutta algorithm. To study the quench dynamics of the edge
solitons, we also employed the Runge-Kutta algorithm for the simulations of
the evolution of the initial excitations in the framework of Eqs. (\ref{eq1}%
)-(\ref{eq3}). A sufficiently small time step was used to ensure the
accuracy of the simulation results. \newline

\noindent \textbf{Acknowledgements}

\noindent R.L., X.K., and W.W. were sponsored by the National Key Research
and Development Program of China (Grant No. 2022YFA1404902), National
Natural Science Foundation of China (Grant No. 12104353), and Fundamental Research Funds for the
Central Universities (Grant No. QTZX25086).
P.L. was sponsored by the National Natural Science Foundation of China (11805141) and
Basic Research Program of Shanxi Provence (202303021211185).
Y.L. was sponsored by the National Natural Science Foundation of China (NSFC) under Grant No.
62271366 and the 111 Project. The work of B.A.M. was supported, in part, by
the Israel Science Foundation through grant No. 1695/22. The numerical
calculations performed in this work were supported by the High-Performance
Computing Platform of Xidian University. \newline

\noindent \textbf{Author Contributions}

\noindent R.L. conceived the idea. R.L., X.K., W.W., and Y.W. performed the
theoretical calculations and simulations. R.L., X.K., W.W., Y.J., and H.T.
designed and conducted the experiments.
R.L., B.A.M., P.L., and W.W. wrote the
manuscript. R.L., B.A.M., and Y.L. supervised the project.
\newline

\noindent \textbf{Data availability}

\noindent The data that support the findings reported in this paper are
available from the corresponding authors upon reasonable request. \newline

\noindent \textbf{Code Availability}

\noindent The program code used in this study is available from the
corresponding authors upon reasonable request. \newline

\noindent \textbf{Competing Interests}

\noindent The authors declare no competing financial interests.\newline

\end{document}